\documentclass[12pt,preprint]{aastex}
\pdfoutput=1
\usepackage{graphicx}
\usepackage{url}
\newcommand{\beq}{\begin{equation}}
\newcommand{\eeq}{\end{equation}}
\newcommand{\beqa}{\begin{eqnarray}}
\newcommand{\eeqa}{\end{eqnarray}}

\def\la{\lower.5ex\hbox{$\; \buildrel < \over \sim \;$}}
\def\ga{\lower.5ex\hbox{$\; \buildrel > \over \sim \;$}}
\begin{document}

\title{Discovery of the hot Big Bang: What happened in 1948}

\author{P.~J.~E. Peebles}  
\affil{Joseph Henry Laboratories, Princeton University, \\ Princeton, NJ 08544}

\begin{abstract}
The idea that the universe is filled with the thermal radiation now termed the Cosmic Microwave Background was first discussed in eleven publications in the year 1948. These papers offer a detailed example of the process of development of a new and now very productive line of research, and of the confusion that can attend new ideas. The confusion in this case left a common misunderstanding of the considerations that motivated the idea of the sea of radiation.
\end{abstract}

\maketitle

\section{Introduction}\label{Sec:Intro}

Measurements of the sea of thermal Cosmic Microwave Background radiation (the CMB) by the WMAP and PLANCK satellites, the ground-based Atacama Cosmology Telescope and South Pole Telescope,\footnote{These projects are discussed in the sites \url{http://map.gsfc.nasa.gov/,
http://www.rssd.esa.int/index.php?project=planck,
http://www.princeton.edu/act/,
http://pole.uchicago.edu/}}
and other instruments have produced remarkably detailed probes of the structure and evolution of the universe and, with the other cosmological tests, make the case for the expanding hot Big Bang cosmology (the relativistic $\Lambda$CDM theory with its cosmological constant and nonbaryonic dark matter) about as compelling as it gets in natural science. (My assessment of the evidence is explained in Peebles 2013.) Kragh (1996), in {\it Cosmology and Controversy}, gives an overview of how we arrived at this cosmology. In {\it Finding the Big Bang}, Peebles, Page, and Partridge (2009) present a more detailed account of what happened in the particularly interesting decade of the 1960s, when circuitous routes led to the realization that space is filled with a sea of radiation, the CMB, and work began on exploring methods of measuring this radiation and analyzing what the measurements might teach us. Also worth special attention is the year 1948, when George Gamow, his student Ralph Asher Alpher, and Robert Herman hit on the argument for a hot Big Bang with its thermal radiation. I offer a reading of what happened in 1948 from what may be gleaned from the unusually dense series of papers on the subject that appeared in that year.

References to the developments in 1948 tend to focus on the paper by Alpher, Bethe, and Gamow (1948), perhaps because it was the first to appear, perhaps in part because of the playful addition of Hans Bethe to the author list. For example, The Astrophysics Data System lists, for the years 2007 through November 2013, 54 citations of the Alpher-Bethe-Gamow paper, 20 citations of Gamow (1948a), which is the first published discussion of the physics of formation of chemical elements and galaxies in a hot Big Bang, and 12 citations of Alpher and Herman (1948a), which presents the first estimate of the temperature of the CMB. Mather~(2007) and Smoot~(2007), in lectures on their Nobel Prizes in recognition of great advances in measurements of the CBM, both refer to the Alpher-Bethe-Gamow paper but not the other two. This is ironic, because Alpher, Bethe, and Gamow (1948) assume a cold Big Bang cosmology, and it makes no mention of thermal radiation. Also, the picture in this paper is physically inconsistent. The inconsistency was resolved later that year by the proposal that elements formed in the early universe in a sea of thermal radiation.

In {\it Genesis of the Big Bang}, Alpher and Herman (2001) recall the Alpher-Bethe-Gamow paper, in the section with that title, but not that there was a change of thinking in 1948 from a cold to hot Big Bang. Kragh (1996), in the section ``Alpha, Beta, Gamma, Delta,'' and Alpher and Herman (1988), in ``Reflections on Early Work on `Big Bang' Cosmology,'' take note of the change but not the reason. This passes over some interesting  considerations that motivated the introduction of ideas that have become central to the established theory of the evolution of the universe. 
 
The development of these considerations in 1948 can be closely studied because there were just three principal actors, and they were publishing rapidly, perhaps even faster than they could reflect on what they were doing, giving us eleven publications in that year (counting Alpher's thesis, a conference report, and a paper and a book that were published the next year but written in 1948 or earlier). This offers a close view of a flow of new ideas that has grown into big science. I mention missteps along the way through 1948. The reader should bear in mind that missteps are a natural part of research in progress, perhaps denser here than usual because the publications were denser than unusual. 

\begin{figure}[htpb]
\begin{center}
\includegraphics[angle=0,width=5.5in]{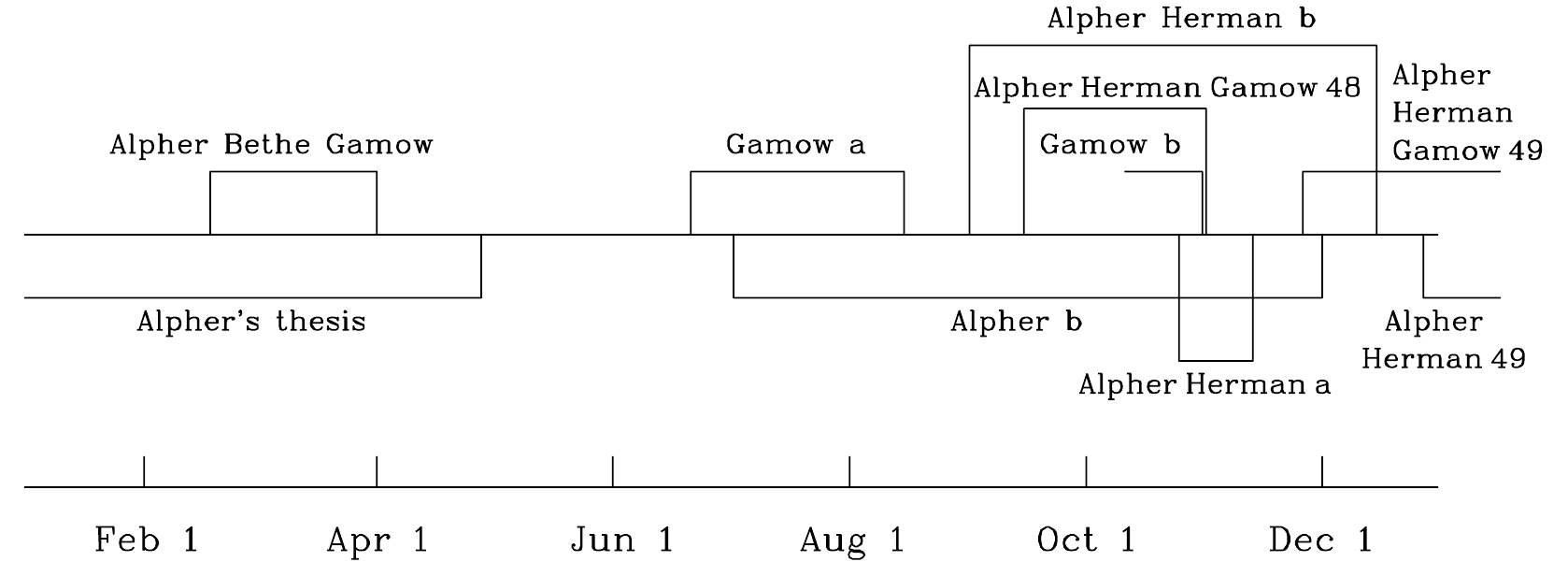} 
\caption{\label{Fig1} \footnotesize  Timeline for 1948 papers on the formation of elements and galaxies. Rectangles span dates of submission and publication. Bent lines mark the dates of presentation of Alpher's thesis, publication of Gamow (1948b), submission of Alpher and Herman (1949), and oral presentation of Alpher, Herman and Gamow (1949).}
\end{center}
\end{figure}

Figure 1 is a timeline for ten of the papers to be discussed. Bent lines mark the date of presentation of Alpher's (1948a) doctoral dissertation, 28 April 1948, the publication of Gamow (1948b) (its date of submission is not recorded), the reception at the journal of Alpher and Herman (1949), in late 1948, and the abstract for a talk at the  meeting of the American Physical Society on November 26 and 27, 1948 (Alpher, Herman, and Gamow 1949). The other papers are marked by the date of reception at the journal and the date of publication. An eleventh publication to consider is the book, {\it Theory of Atomic Nucleus and Nuclear Energy-Sources} (Gamow and Critchfield 1949). The preface is dated September 1947, the text mentions Alpher's thesis work in progress but predates Alpher, Bethe, and Gamow (1948), and an appendix on ``New important findings, which appeared during the printing of this book'' postdates Gamow (1948b).

Section \ref{Sec:bb} reviews relevant ideas of the hot Big Bang cosmology. I begin Section~\ref{Sec:publs} with a discussion of  Alpher's thesis, because we can be reasonably sure work on the thesis was well advanced when the paper Alpher, Bethe, and Gamow (1948) was submitted. This is followed by examinations of the other papers roughly in the order of the timeline. The story this reveals is not straightforward; we see the confusion of exploration of new ideas. Lessons from this history are offered in Section~\ref{Sec:unexamined}. My conclusions about who likely was responsible for the major ideas and what led to them are explained in Section~\ref{Sec:concl}. 

\section{The relativistic hot Big Bang cosmological model}\label{Sec:bb}

The evolution of a near homogeneous universe is measured by its expansion parameter, $a(t)$, where $t$ is the time measured by an observer moving with the matter, and the mean distance between conserved particles such as baryons varies in proportion to $a(t)$, meaning the number density of conserved baryons varies as $a(t)^{-3}$. The homogeneous expansion of the universe causes the temperature of a homogeneous sea of thermal radiation to decrease as $T\propto a(t)^{-1}$, while preserving the thermal spectrum (as shown by Tolman 1934, p. 428).  (In more detail,  the interaction with matter, which tends to cool more rapidly than radiation, is minor in conditions of interest because the radiation heat capacity is so large. The addition to the radiation entropy when the temperature fell low enough to allow annihilation of the early thermal sea of electron-positron pairs increased the radiation temperature by a modest factor without disturbing the spectrum, because thermal relaxation was fast then.) The mass densities in radiation and nonrelativistic matter scale as 
\beq
\rho_{\rm rad}\propto T^4\propto a(t)^{-4}, \qquad \rho_{\rm mat}\propto a(t)^{-3}. \label{eq:densityvsa}
\eeq
In the standard hot Big Bang cosmology the mass density in thermal radiation is much larger than in matter during the time of early element formation, and the more rapid decrease of $\rho_{\rm rad}$ with increasing $a(t)$ makes the mass density  in matter larger than radiation well before the present epoch. 

The thermal radiation plays two important roles in early element formation. First, the energy density in radiation determines the rate of cosmic expansion (which scales as the square root of the total mass density). Second, as the universe expands and cools the thermal radiation sets the critical temperature, close to $T_i\simeq 10^9$\,K,  at the start of element buildup. This is determined by the changing balance of rates of radiative capture of neutrons (n) by protons (p) to form deuterons (d), and dissociation of deuterons by photons ($\gamma$),
\beq
{\rm n}+{\rm p}\leftrightarrow {\rm d}+\gamma. \label{eq:d}
\eeq
At temperature slightly above $T_i$ the reaction toward the left, photodissociation, strongly suppresses the number density of deuterons. As the universe expands and  the temperature falls slightly below $T_i$ the balance shifts to favor deuterons, because the fraction of thermal photons energetic enough for photodissociation finally is too small to compete with the formation of deuterons. (In the dilute gas of nucleons in the early universe this critical temperature is a slow logarithmic function of the nucleon number density. The early work assumes neutrons were at least as abundant as protons at $T_i$. Hayashi 1950 shows how the weak interactions determine the relative abundance of neutrons and protons, and Alpher, Follin, and Herman 1953 present a detailed computation of the n/p ratio.) Deuteron formation by the reaction~(\ref{eq:d}) sets the threshold for element buildup because once deuterons can accumulate the conversion of deuterons to heavier elements by particle exchange reactions such as ${\rm d}+{\rm d}\rightarrow {\rm t} + {\rm p}$ is much faster than production of tritium (t) or anything heavier out of interactions of free neutrons and protons. 

The condition that a reasonable amount of deuterons can accumulate, to be burned to heavier elements, sets the wanted value of the baryon number density  $n_i$ at the radiation temperature $T_i$ at the onset of  element formation. The value of $n_i$ at $T_i$ with the relations in equation~(\ref{eq:densityvsa}) determines the expected present temperature $T_f$ of the radiation, the CMB, given the present baryon number density $n_f$, 
\beq
T_f = T_i(n_f/n_i)^{1/3}. \label{eq:presentTemp}
\eeq
Application of this relation is an early start to the network of cosmological tests (Planck Collaboration XIV 2013). 

Gamow took an important step toward the developments in 1948 by pointing out that if the universe expanded from a state of very large mass density --- commonly termed the Big Bang --- then according to general relativity theory the early universe had to have expanded very rapidly. Gamow (1946a) showed a simple way to understand this. Consider a sphere of radius $l$ that contains matter with density $\rho$. The sphere contains mass $M\sim\rho l^3$, ignoring numerical factors, so according to Newtonian gravity the potential energy at the surface is 
\beq
GM/l\sim G\rho l^2\sim v^2\sim (l/t)^2,\label{eq:Grho}
\eeq
where $G$ is Newton's constant and $v\sim l/t$ is the escape velocity, with $t$ the expansion time. Canceling the radius $l$ from the second and last expressions leaves $t\sim (G\rho)^{-1/2}$ for the time $t$ of expansion to mass density $\rho$ in the early universe. This relation also applies in general relativity, with a numerical factor that depends on whether the mass density in radiation is important. Gamow's point is that the very large values of $\rho$ early on had to have been accompanied by very small expansion times $t$, meaning one should consider how element abundances might be determined by rapid non-equilibrium processes. 

In the paper Gamow (1946a) the proposal is that 
\begin{quotation}\noindent
if free neutrons were present in large quantities in the beginning of the expansion, the mean density and temperature of expanding matter must have dropped to comparatively low values {\it before} these neutrons had time to turn into protons. We can anticipate that neutrons forming this comparatively cold cloud were gradually coagulating into larger and larger neutral complexes which later turned into various atomic species by subsequent processes of $\beta$-emission.
\end{quotation}
Gamow takes note of the earlier idea that element abundances were set by near thermal equilibrium in the early universe, at temperatures that would have been on the order of $10^{10}$\,K (Chandrasekhar and Henrich 1942). He rejects this idea because there is not enough time for thermal relaxation. The thought that radiation at a similar temperature might accompany element formation in a non-equilibrium process appears only in 1948. But Gamow (1946a) is a good start.  

\section{The publications}\label{Sec:publs}

\subsection{Alpher's doctoral dissertation}\label{Sec:Alpher}

The thesis Alpher (1948a) presents an analysis of the idea Alpher attributes to Gamow (1946a), that ``the elements must have been formed by some rapid continuous building-up process involving successive neutron captures.''  The analysis requires radiative capture cross sections $\sigma$ of neutrons by the atomic nuclei. Kragh (1996) reports that Alpher attended a talk by Donald J. Hughes (1946a) at a meeting of the American Physical Society (June 20-22, 1946) on capture cross section measurements, which led him to  Hughes (1946b), a declassified document on measurements that used ``1-Mev pile neutrons.'' This is Alpher's most important source of data on neutron capture cross sections.  At energy $\sim 1$\,MeV the cross sections for many elements vary inversely as the relative velocity $v$, so one may usefully tabulate $\sigma v$ or $\sigma\sqrt{E}$, where $E$ the kinetic energy. Hughes (1946b) reports that the mean trend with atomic weight $j$ at $j<100$ is
\beq
\sigma_j\sqrt{E} = 10^{0.03j - 1.00}, \label{eq:Hughessv}
\eeq 
where the units are electron volts and barns (and 1~barn $= 10^{-24}$~cm$^2$). 

The rate of change of  abundances of the atomic nuclei as as a function of atomic weight in the process of element buildup by neutron capture is expressed in Alpher's thesis as 
\beq
{dn_i\over dt} = f(t)(\lambda_{i - 1}n_{i - 1} - \lambda_in_i). \label{eq:rates}
\eeq 
The first term on the right-hand side represents the rate of production of nuclei with atomic weight $i$  by neutron capture by nuclei with atomic weight $i-1$, and the second term is the rate of loss by promotion to $i+1$. The $\lambda_i$ are proportional to the neutron capture cross section $\sigma_i$, and $f(t)$ takes account of the decay of the neutrons and the evolution of the particle number densities as the universe expands. The same relation expressed in terms of the $\sigma_i$  instead of $\lambda_i$ is in Alpher, Bethe, and Gamow (1948), and in text that predates that in Gamow and Critchfield (1949).  A footnote in the latter is ``More detailed calculations on this non-equilibrium process are being carried out by R. A. Alpher''. 

Numerical solutions to equation~(\ref{eq:rates}) were a daunting challenge for computation at that time, so Alpher simplified the problem by ignoring the time-dependence due to the decay of neutrons and expansion of the universe, parametrizing the exposure to neutrons as the integral $\int_{t_0}^{t_1}n\,dt$ of the baryon number density over the time of element buildup, $t_0$ to $t_1$. This assumes, as noted in Alpher, Bethe, and Gamow (1948), that buildup commences at time $t_0$ after the start of expansion, because the integral diverges at $t_0\rightarrow0$ and $n\rightarrow\infty$.

The conclusion in Alpher (1948a) is that the general trend of $\sigma_iv$ with atomic weight $i$ can fit the observed trend with $i$ of  element abundances if
\beq
\int_{t_0}^{t_1} n(t)\,dt \equiv \langle n t \rangle = 0.81\times 10^{18} \hbox{ s cm}^{-3}. \label{eq:Ant}
\eeq 
Alpher's computed relative abundances are a distinct improvement over the preliminary estimates in Gamow and Critchfield (1949, Fig.~59). 

The value of Alpher's $\langle n t \rangle$ is of particular interest for this story because it can be compared to the relation between density and time in the cosmological model, so it is worth pausing to consider a simple order-of-magnitude check. Equation~(\ref{eq:Hughessv}) at $j=1$, which is about appropriate for formation of deuterons, gives
\beq
\sigma v \simeq 1.5\times 10^{-19}\hbox{ cm}^3\hbox{ s}^{-1}. \label{eq:sigv}
\eeq
The condition for significant but not excessive accumulation of deuterons and heavier elements in a roughly equal mix of neutrons and protons reacting at expansion time $t$ is, following Gamow (1948a),
\beq
\sigma v n t\sim 1, \label{eq:svnt}
\eeq
meaning the mean free time for ${\rm n}+{\rm p}\rightarrow {\rm d}+\gamma$ is comparable to the expansion time $t$.  With equation~(\ref{eq:sigv}), this condition indicates
\beq
\langle n t \rangle \sim (\sigma v)^{-1}\sim 7\times 10^{18} \hbox{ s cm}^{-3}. \label{eq:mynt}
\eeq 
Equations~(\ref{eq:Ant}) and~(\ref{eq:mynt}) are reasonably close, considering that the latter is a crude though  straightforward estimate. That is, Alpher's condition on $\langle n t \rangle$ is secure within the assumptions. Consider now the problem Alpher points out in reconciling $\langle n t \rangle$ with what would be expected from the cosmological model. 

Alpher (1948a) considers the assumption that the mass in the early universe is dominated by baryonic matter with mass density $\rho_{\rm mat} = nm$, where $m$ is the nucleon mass. In this model the expansion time $t$ at density $\rho_{{\rm mat}}$ computed from much higher density satisfies
\beq
\rho_{{\rm mat}} = n m = (6\pi Gt^2)^{-1}, \hbox{ or }  n t= 5\times 10^{29}(\hbox{s}/t) \hbox{ s cm}^{-3}, \label{eq:cosnt}
\eeq
for $t$ in seconds. If this value of $nt$ were to be comparable to $ \langle n t \rangle$ in equation~(\ref{eq:Ant}) or~(\ref{eq:mynt}) nucleosynthesis would have to started when the expansion time was
\beq
t_i\sim 10^{11}\hbox{ s}. \label{eq:vastinconsistency} 
\eeq
But this is many orders of magnitude longer than the neutron half-life, $t_{1/2}\sim 1000$\,s. In another way to put it, at expansion time $t$ equal to $t_{1/2}$ equations~(\ref{eq:sigv}) and~(\ref{eq:cosnt}) indicate that $\sigma v n t_{1/2}\sim 10^8$, meaning there is ample time for neutrons to combine with all the protons as they appear from neutron decay, leaving no hydrogen, an absurd situation. If this early element buildup were somehow prevented until the time $t_i$ in equation~(\ref{eq:vastinconsistency}) then the neutrons created with the expanding universe would have long since decayed. What could have produced a flood of neutrons this late in the game? There are ways out, but none looked attractive then or now. 

The  established resolution to Alpher's inconsistency is that the early universe was hot, with mass density dominated by a near homogeneous sea of thermal radiation. There are steps toward this resolution in Alpher's thesis. Worth noting first is his discussion of the likely temperature of the neutrons:
\begin{quotation}
\noindent It seems reasonable to suppose that the temperatures during the process were well above the resonance regions of the elements. On the other hand, temperatures of an order of magnitude higher than 1~Mev correspond to neutron energies larger on average than the binding energy per particle in nuclei. A temperature of 10~Mev, therefore, must be well above the temperatures during the period of formation. A temperature above $10^3$~ev, and less than 10~Mev, perhaps of the order of $10^5$~ev (about $10^9\,^\circ$K) appears to be approximately the correct one.
\end{quotation} 
Alpher's lower bound is from the condition that the neutrons must be moving fast enough to avoid the large  neutron capture resonances that would upset the relation between the trends of relative abundances and capture cross sections. Alpher's choice of $T_i\simeq 10^9$\,K is the geometric mean of his two bounds, perhaps meaning it is just a convenient round number. Or it may be significant that Alpher's $T_i$ agrees with the later appearance of Gamow's (1948a) value of the temperature at the start of element buildup in a hot Big Bang. In this connection it is to be noted that both of Alpher's bounds refer to neutron kinetic energies. Equilibrium with a sea of thermal radiation is a separate condition, perhaps plausible, but to be based on a picture of what happened earlier in the expansion that might have produced relaxation to thermal equilibrium. It is not clear whether Alpher (1948a) is making this distinction in writing that ``it may be suggested that an appreciable amount of energy is present in the form of radiation'' and 
\begin{quotation}
\noindent we have seen that the temperature obtaining during the process must have been of the order of $10^5$~ev $\simeq 10^9\, ^\circ\hbox{K}$. But the density of blackbody radiation at this temperature would have been 
$$
\rho_{\rm radiation} = 0.848\times 10^{-35}T^4\simeq 10\hbox{ gm}/\hbox{cm}^3.\label{eq:rhorad}
$$
This is many orders of magnitude greater than the density of matter given by the particular cosmological model used. It would appear, then, that radiation was dominant in determining the behaviour of the universe in the early stages of its expansion, and the cosmological model which we have used is not correct. An  interpretation of the starting time and initial density for the neutron-capture process will therefore require the development of a new cosmological model. 
\end{quotation} 
Since Alpher neglected the time-evolution of the factor $f$ in finding numerical solutions to the buildup equation~(\ref{eq:rates}), the change to a new cosmological model would not affect Alpher's condition in equation~(\ref{eq:Ant}). 

Alpher's remarks about thermal radiation are an important step toward the new cosmology, but not yet a suggestion of how the radiation would allow a consistent theory of element buildup. That appeared later in the year. 

\subsection{Alpher, Bethe, and Gamow}  \label{Sec:ABG}

The paper Alpher, Bethe, and Gamow (1948) was submitted to The Physical Review two months before Alpher presented his thesis (Fig.~\ref{Fig1}). It notes that ``More detailed studies of Eqs. (1) leading to the observed abundance curve and discussion of further consequences will be published by one of us (R. A. Alpher) in due course.'' (The reference is to eq.~[\ref{eq:rates}] above.) The paper does not mention the serious mismatch (eqs.~[\ref{eq:Ant}] and~[\ref{eq:cosnt}]) pointed out in the thesis, and it does not mention the idea of a sea of thermal radiation.  

Alpher, Bethe, and Gamow (1948) state that the fit of element buildup to observed abundances requires that ``the integral of $\rho_n\,dt$ during the build-up period is equal to $5\times 10^4$ g~sec./cm$^3$.'' In the notation of equation~(\ref{eq:Ant}) this is
\beq
\langle nt\rangle \sim 3\times 10^{28}\hbox{ s cm}^{-3}. \label{eq:ABGnt}
\eeq
In the matter-dominated cosmology assumed in this paper the nucleon number density varies with time $t$ as $n\propto t^{-2}$. Since the integrated neutron exposure $\int n\,dt$ thus diverges at $t\rightarrow 0$ the paper assumes element buildup commenced at time $t_0$, perhaps because the neutrons were too hot to be captured before that, or perhaps because the expanding universe bounced from earlier collapse at about the density corresponding to $t_0$. Alpher, Bethe, and Gamow (1948) point out that $nt$ from the cosmological model (eq~[\ref{eq:cosnt}]) agrees with equation~(\ref{eq:ABGnt}) if nucleosynthesis commenced at $t_0\simeq 20\,s$. This is less than the neutron half-life $t_{1/2}$, a not unreasonable situation. Something is very wrong, however, because equation~(\ref{eq:ABGnt}) is ten orders of magnitude larger than the value in Alpher's thesis (eq.~[\ref{eq:Ant}], confirmed by the rough check in eq~[\ref{eq:mynt}]). 

Gamow (1948a) writes that the time integral of $\rho_{\rm mat}$ ``was given incorrectly in the previous paper$^2$ because of a numerical error in the calculations.''  (The reference is to Alpher, Bethe, and Gamow 1948). What might be the  nature of the numerical error? Kragh (1996) attributes it to ``a trivial error of calculation where a factor of $10^4$ had been used instead of the correct $10^{-4}$.'' But there may be more to the story,  because the erroneous value of $\langle nt\rangle$ allows apparent consistency with the cosmology if element buildup started at $t_0\sim 20\hbox{ s}<t_{1/2}$. One might imagine that, two months before completion of his thesis, Alpher had not yet recognized the mismatch between nucleosynthesis and expansion rates. If so one might also imagine that Alpher made an error in copying $\langle n t\rangle$ to the Alpher, Bethe, and Gamow paper, one that happens to obscure the mismatch. Or maybe Gamow was unwilling to acknowledge the inconsistency until he found how to resolve it. 

The first proposal of a resolution of sorts postulates a bounce at maximum density corresponding to $t_0$ in a rotating universe (Gamow 1946b). But the idea seems contrived: Where would the neutrons have come from?  (And the singularity theorems now convincingly argue against a bounce under standard general relativity theory.) Gamow (1948a) offers a simpler way. 

\subsection{Gamow}

\subsubsection{Element buildup}\label{Sec:Element buildup}

As noted in Section~\ref{Sec:bb}, the established theory is that element buildup starts with deuterons.  Gamow (1948a) writes
\begin{quotation}
Since the building-up process must have started with the formation of deuterons from the primordial neutrons and the protons into which some of these neutrons have decayed, we conclude that the temperature at that time must have been of the order $T_0\simeq 10^9\,^\circ$K (which corresponds to the dissociation energy of deuterium nuclei), so that the density of radiation $\sigma T^4/c^2$ was of the order of magnitude of water density.
\end{quotation}
Gamow (1948b) refers to Bethe (1947) for the cross section for ${\rm n} + {\rm p} \rightarrow {\rm d} + \gamma$,
which in his 1948a paper he states is $\sigma\simeq 10^{-29}$\,cm$^2$. At $T_i=10^9$\,K this is
\beq
\sigma v \sim  10^{-20}\hbox{ cm}^3\hbox{ s}^{-1}. \label{eq:Gsigv}
\eeq
The mass density $\rho_{{\rm rad}}$ in radiation is 
\beq
\rho_{\rm rad} = a_{\rm B}T^4/c^2 = 8.40\,(T/10^9\,{\rm K})^4\hbox{ g cm}^{-3},\label{eq:rho_rad}
\eeq
where $a_{\rm B}$ is Stefan's constant, and, if $\rho_{\rm rad}$ is much larger than the mass in matter, the expansion time $t$ to temperature $T$ is
\beq
t  =\sqrt{3/32\pi G \rho_{\rm rad}} = 231\,(10^9\,{\rm K}/T)^2\hbox{ s}. \label{eq:t_rad}
\eeq
Gamow's (1948a) approximate condition $\sigma v n t\sim 1$ at the start of element buildup (as in eq.~[\ref{eq:svnt}]), with $t_i$ at $T_i=10^9$\,K from equation~(\ref{eq:t_rad}) and $\sigma v$ from equation~(\ref{eq:Gsigv}), gives Gamow's (1948a) estimate of the baryon number density at the start of element buildup,  
\beq
n_i\sim (\sigma v\, t_i)^{-1} \sim 10^{18}\hbox{ cm}^{-3} \hbox{ at } T_i=10^9\hbox{ K}. \label{eq:Gni}
\eeq

Another way to state the condition $\sigma v n t\sim 1$ is that in an equal mix of neutrons and protons the mean free time of a neutron is comparable to the expansion time, meaning that about half the nucleons would be captured in heavier nuclei, leaving about 50\% hydrogen by mass. This is illustrated in more detail in Figure~2 in Gamow (1948b), which shows a numerical solution for the buildup of deuterons, ignoring the reactions that would convert  deuterons to heavier elements. As expected, the numerical solution under Gamow's condition approaches comparable masses in hydrogen and heavier elements. Gamow (1948b) was of the opinion that ``hydrogen is known to form about 50 percent of all matter.'' That is somewhat less hydrogen than the Hall Harrison (1948) analysis of the composition of the Sun, about 70\% hydrogen, 30\% helium, and a few percent heavier elements, with similar values in Schwarzschild (1946) that are quoted in Gamow and Critchfield (1949, pp. 282-283). Hoyle (1950) expresses doubt about the helium abundance, but the large mass fraction in helium that Gamow favored is not far from present constraints. 

To fit these Solar abundances, element buildup would have to have converted most of the deuterons shown in Gamow's (1948b) Figure~2 to helium, while allowing only a small fraction to have grown more massive than that, because only a few percent of the cosmic mass is in heavier elements. Gamow (1948a,b) does not comment on this condition. The role of Alpher's mass-5 gap in meeting the condition is discussed in Section~\ref{Sec:AlpherPaper}. 

Gamow's value of the temperature at the start of element buildup is important for this story, so it is worth pausing to consider what he meant by the ``dissociation energy of deuterons.'' As discussed in Section~\ref{Sec:bb}, it is now established that element buildup commenced when the temperature fell to $T_i\simeq 10^9$\,K, at which point  there were no longer enough photons energetic enough for photodissociation of deuterons to keep up with deuteron formation. This temperature is well approximated by the Saha (detailed balance) expression for relative abundances at thermal equilibrium. It corresponds to characteristic energy $kT_i\simeq 0.1$\,MeV, which well below the deuteron binding energy, $B=2.2$\,MeV.  So what did Gamow (1948a) mean by the ``dissociation energy''? The deuteron binding energy is given and analyzed in Gamow and Critchfield (1949, pp. 25 and 43), in text written before Gamow (1948a). The Saha expression is discussed in Gamow and Critchfield (pp. 309 - 310), who note that interesting relative abundances of heavier elements are produced at Saha equilibrium at temperatures approaching $10^{10}$\,K  (quoting from Chandrasekhar and Henrich 1942). The value of $T_i$ for deuteron buildup is lower, about $10^9$~K, because the binding energy per nucleon is smaller.  The evidence is then that Gamow knew the Saha expression, and its application to deuterons, and that by dissociation energy he meant the threshold temperature $T_i\simeq 10^9$\,K for deuteron buildup, not the larger deuteron binding energy. But since I have not found a statement by Gamow to this effect it might be considered a plausible conjecture. 

I conclude that Gamow (1948a,b) presents two important points. First, in a radiation-dominated universe the onset of element buildup is set by the threshold for thermal dissociation of deuterons. Second, the rates of nuclear reaction and cosmic expansion can be reconciled by the adjustment of $n_i$. In a matter-dominated universe the choice of $n_i$ at some arranged start of element buildup determines both the reaction rate and the expansion time (eq.~[\ref{eq:cosnt}]), so special arrangements are required both to suppress early buildup and to reconcile rates of neutron decay and capture with the rate of cosmic expansion when buildup does start. Resolution of these problems by a hot Big Bang does not make it so, of course, but there are tests, beginning with the translation of $n_i$ in equation~(\ref{eq:Gni}) to the expected present radiation temperature (eq.~[\ref{eq:presentTemp}]), which can be compared to the measured temperature of the CMB. Gamow's reconciliation by the recognition of the effects of radiation is memorable because it is now seen to pass an abundance of tests. 

\subsubsection{Galaxy formation}

The title of Gamow's (1948a) paper, {\it The Origin of Elements and the Separation of Galaxies}, shows he was also thinking about the role of the thermal radiation in galaxy formation after element formation. In Alpher and Herman (2001), Alpher recalls Gamow telling him about Lifshitz's (1946) analysis of the gravitational evolution of departures from homogeneity in an expanding universe, just when Alpher had started writing up his work Gamow had assigned him on this problem. The analysis shows that in a universe with the equation of state of radiation, $p=\rho c^2/3$, and in linear perturbation theory, gravity cannot cause the growth of density fluctuations on scales less than the Hubble length (the distance at which the redshift-distance relation extrapolates to the velocity of light). But Lifshitz's analysis does not include the situation of interest here, the gravitational growth of concentrations of low-pressure matter, baryons, when the mass density may be dominated by radiation (and assuming  negligible drag on the matter by the radiation). Here, as Gamow argued, perhaps by intuition, small irregularities in the matter distribution cannot grow into galaxies unless the mass density in radiation is less than about the mass density in matter. Since the ratio of mass densities in radiation and matter varies with time in proportion to  the radiation temperature (eq.~[\ref{eq:densityvsa}]), one sees that the mass densities in radiation and matter at $10^9$\,K in equations~(\ref{eq:rho_rad}) and~(\ref{eq:Gni}) translate to
\beq
\rho_{\rm eq}\sim 10^{-26}\hbox{ g cm}^{-3}, \quad T_{\rm eq} \sim 200\hbox{ K}, \label{eq:GTeq}
\eeq 
at equality of mass densities in matter and radiation. Gamow proposed that this marks the start of the growth of mass concentrations such as galaxies. He also realized that the matter density and temperature, which he took to be the same as the radiation temperature, determine the time-independent Jeans mass of the smallest gravitationally-assembled clouds of baryons. (Gamow's 1946a numbers are not very close to eq.~[\ref{eq:GTeq}], but Gamow seems not to have cared about such details.)

It might be noted that the large overestimate of Hubble's constant at that time, with the use of a close to realistic present matter density, implied   that the mass density at $T_{\rm eq}$ would have been subdominant to space curvature or the cosmological constant (as discussed in Sec.~\ref{Sec:AH}). The apparently low mass density at $T_{\rm eq}$ would have meant that the expansion rate at $T_{\rm eq}$ is so rapid that galaxies could not form at all. The point was made, however, that thermal radiation plays an important role in cosmic structure formation. The modern theory adds many details, but this is a start. 

\subsection{Alpher's paper}\label{Sec:AlpherPaper}

The paper Alpher (1948b) is the version of his thesis published in {\it The Physical Review}  (a standard practice,  for although the thesis is copyrighted and publicly available access to the journal is much easier). There are two interesting additions. First, Alpher (1948b) points out that 
\begin{quotation}
\noindent it appears to be impossible to build past atomic weights 5, 8 and 11 by a process of neutron captures alone, because of the absence of stable atomic nuclei at these atomic weights. However, because of the high temperature, reactions with deuterons or with tritons may be reasonably introduced for the low Z nuclei, and thereby make it possible to  bridge these gaps.
\end{quotation}
Alpher and Herman (2001) recall that a colloquium by Alpher in late 1949 led Enrico Fermi and Anthony Turkevich to explore the reactions leading from deuterons to heavier elements. Alpher and Herman (1950) report that Fermi and Turkevich found no plausible reactions that yield more than exceedingly small production of elements heavier than helium. The Fermi and Turkevich numerical integration of the formation and burning of deuterons yields relative abundances of hydrogen, deuterium, and the two isotopes of helium that follow the general pattern of present data, though there is too much remnant deuterium because their baryon density is well below current constraints. Alpher's mass-5 gap proves to be effective, and relevant, because as noted in Section~\ref{Sec:Element buildup} the buildup must have produced much more helium than heavier elements. It is now persuasively established that the bulk of the heavier elements were produced in stars, in part by fast neutron capture in a process similar to the 1948 picture. 

The second notable addition, which appears just after Alpher's discussion of the problem with element formation in a matter-dominated cosmology, is the paragraph
\begin{quotation}
Preliminary calculations of a cosmological model involving black body radiation only$^{24}$ (the effect of matter on the behavior of the model being negligible in the early stages because of the great difference between radiation and matter density) indicate that at some 200 to 300 seconds after the expansion began the temperature would have dropped to about $10^9\,^\circ$K, at which time the neutron-capture could have begun.
\end{quotation}
(The reference is to Tolman 1931.) Alpher (1948b) does not say whether the temperature mentioned here is based on the considerations in his thesis, or on Gamow's (1948a) dissociation energy. There is no mention of  Gamow's (1948a) argument about how thermal radiation could relieve the problem with a matter-dominated cosmology. Alpher's  paper was received by the journal 11 days after receipt of Gamow (1948a). Perhaps Alpher did not know about Gamow's argument. Perhaps he knew but did not want to refer to a recently submitted paper. Or perhaps, as discussed next, Alpher had come to distrust Gamow's use of the thermal threshold condition. 

\subsection{The Alpher and Herman papers} \label{Sec:AH}

The papers Alpher (1948a,b), Alpher and Herman (1948b), Alpher, Herman, and Gamow (1948), and the later review Alpher and Herman (1950), mainly report progress in the challenging problem of numerical computation of   element buildup by neutron and proton capture. These papers could be mined for another history, on the coevolution of computers and the numerical methods of using them. They also offer indications of the thinking about how to estimate physical conditions for element formation in the early stages of expansion of a hot Big Bang. 

The memorable statement in the short paper Alpher and Herman (1948a) is that ``the temperature in the universe at the present time is found to be about $5^\circ$\,K.'' This is famously close to what was measured beginning more than a decade later,  $2.726$\,K. However, there are some issues about how Alpher and Herman arrived at their temperature.

A computation of the expected present radiation temperature requires the mass density in matter (baryons of course) wanted for acceptable element formation  in the early universe. Estimates of this matter density are expressed in the 1948 papers as 
\beq
\rho_{\rm mat} = C\, ({\bf s}/t)^{3/2} \hbox{ g cm}^{-3}, \label{eq:Gbrhom}
\eeq
where $C$ is a constant, $t$ is the time of expansion from much larger density, and the time-dependence is that of a radiation-dominated universe. Gamow's (1948b) estimate is 
\beq
C_{\rm Gamow}=7.2\times 10^{-3}. \label{eq:CGamow}
\eeq
It is a little smaller, $C=6\times 10^{-3}$, in Gamow (1948a). Gamow's (1948b) value is stated also in Alpher, Herman, and Gamow (1948), which discusses the possible role of proton reactions, and in Appendix VI in Gamow and Critchfield (1949). 

The paper Alpher and Herman (1948a) lists four corrections to Gamow's computation of the onset of element buildup. (The paper refers to Gamow 1948b, which was published a few days after reception by the journal of the Alpher and Herman 1948a paper, but the general remarks apply as well to Gamow~1948a.) Following the list of four corrections to Gamow (1948b) is the statement that, ``correcting for these errors, we find''
\beq
C_{\rm AH48a}=4.83\times 10^{-4}. \label{eq:AandHrhom}
\eeq

Alpher and Herman (1948b) report solutions to the buildup equations~(\ref{eq:rates}) using the smooth fit to capture cross sections in equation~(\ref{eq:Hughessv}), neglecting the expansion of the universe but taking account of neutron decay (meaning the time scale for element buildup is set by the neutron half-life). Their fit to the broad range of measured relative abundances indicates neutron number density $n_0=3.23\times 10^{15}\hbox{ cm}^{-3}$ at the start of buildup, at which time a ``temperature of about $10^5$~ev$\simeq10^9\,^\circ\,$K is suggested$^4$.'' (The reference is to Alpher 1948b). As noted in Sec.~\ref{Sec:AlpherPaper}, the provenance of this temperature is unclear. Alpher and Herman (1949) offer a correction to this density to take account of the general expansion of the universe, which brings the matter density to 
\beq
C_{\rm AH49} = 1.70\times 10^{-2},
\eeq 
with the caution that this is a rough estimate.

Translation of the early matter density to an estimate of the present radiation temperature requires the present matter density. This density is not mentioned in Alpher and Herman (1948a), but when stated, as in Gamow (1946a), Gamow and Critchfield (1949), Alpher (1948a, p. 18) and Alpher and Herman (1949, 1950), it is consistently 
\beq
\rho_{{\rm mat},f}=10^{-30}\hbox{ g cm}^{-3}. \label{eq:presentdensity} 
\eeq 
The relation between time and temperature in equation~(\ref{eq:t_rad}) brings $C$ to the matter density $\rho_{{\rm mat},i}$ at $T_i=10^9$\,K, and equation~(\ref{eq:presentTemp}) with $\rho_{{\rm mat},i}$ and $\rho_{{\rm mat},f}$ gives the present temperature, $T_f$. The results are  
\beqa
\rho_{{\rm mat},i}&=& 2.0\times 10^{-6}\hbox{ g~cm}^{-3}, \ T_f = 8\hbox{ K}\ \hbox{ for }\ C_{\rm Gamow}, \nonumber \\
\rho_{{\rm mat},i}&=& 1.4\times 10^{-7}\hbox{ g~cm}^{-3}, \ T_f = 19\hbox{ K}\ \hbox{ for }\ C_{\rm AH48a}, \nonumber  \\
\rho_{{\rm mat},i}&=& 4.8\times 10^{-6}\hbox{ g~cm}^{-3}, \ T_f = 5.9\hbox{ K}\ \hbox{ for }\ C_{\rm AH49}. \label{eq:Tf}
\eeqa
The value of $C_{\rm AH48a}$ presented in Alpher and Herman (1948a) makes $T_f$ much larger than the temperature, ``about $5\,^\circ$K'', appearing in this paper, while $C_{\rm AH49}$ brings $T_f$ much closer. Consistent with this, Alpher and Herman (1949) state that their density, $C_{\rm AH49}$, ``corresponds to a temperature now of the order of $5\,^\circ$K,'' the same as  in Alpher and Herman (1948a).

There are two more checks that Alpher and Herman (1948a) use matter density close to $C_{\rm AH49}$ for their estimate of $T_f$, not the value $C_{\rm AH48a}$ in that paper. The first is the Alpher and Herman (1948a) value of the time at equality of the mass densities in matter and radiation. This requires Hubble's constant. Alpher (1948a,b) uses $H_o=1.8\times 10^{-17}$~s$^{-1}$ (or $H_o = 560$ km~s$^{-1}$~Mpc$^{-1}$), for which the critical density is 
\beq
\rho_c=3H_o^2/8\pi G= 5.8\times  10^{-28}\hbox{ g cm}^{-3},
\eeq
the density parameter  in matter is $\Omega_m=\rho_{{\rm mat},f}/\rho_c=0.0017$, and, at present temperature $T_f=5.9$\,K from $C_{\rm AH49}$, the density parameter in radiation is $\Omega_r=a_{\rm B}T_f^4/\rho_cc^2=1.8\times 10^{-5}$. Alpher and Herman (1948a) use the Friedmann-Lema\^\i tre expansion rate equation
\beq
(\dot a/a)^2 = H_o^2\left[\Omega_m(1+z)^3+ \Omega_r(1+z)^4+ \Omega_k(1+z)^2 \right],
\label{eq:FL}
\eeq
where $z$ is the redshift and $\Omega_k=1-\Omega_m-\Omega_r$ is the space curvature term. With these numbers the dominant term in equation~(\ref{eq:FL}) is space curvature back to redshift $1+z_k\simeq\sqrt{\Omega_k/\Omega_r}\simeq 240$, prior to which the expansion rate is dominated by the energy density in radiation. The mass densities in matter and radiation are equal at $1+z_{\rm eq}=\Omega_m/\Omega_r = 100$. At $z_{\rm eq}$ the model expansion rate is dominated by space curvature, so the expansion parameter is increasing approximately linearly with time, and the expansion time at equality is then $t_{\rm eq}\sim H_o^{-1}/(1+z_{\rm eq})\sim 6\times 10^{14}$\,s. This is in reasonable agreement with Alpher and Herman (1948a),  who give $t_{\rm eq}=3.5\times 10^{14}$\,s. The second check is the conclusion that ``The temperature of the gas at the time of condensation was $600^\circ$\,K''. The time of condensation is not defined, but because it appears directly after the discussion of $t_{\rm eq}$ it seems reasonable to follow Gamow in taking it to be $t_{\rm eq}$. If so, then since the matter and radiation mass densities vary as $\rho_{{\rm mat}}\propto T^3$ and $\rho_{{\rm rad}}\propto T^4$, equation~(\ref{eq:rho_rad}) and the third line of equations~(\ref{eq:Tf}) indicate that $T_{\rm eq}=580$\,K, again consistent with Alpher and Herman (1948a). In contrast, the density presented in Alpher and Herman (1948a) (the second line of eqs.~[\ref{eq:Tf}]) gives $T_{\rm eq}=20$\,K, much smaller than they report, again supporting the argument that they were really using a larger density than $C_{\rm AH48a}$, much closer to $C_{\rm AH49}$. One might have hoped for yet another check from the statement in the first paragraph in Alpher and Herman (1948a) that ``the intersection point $\rho_{\rm mat.}=\rho_{\rm rad.}$ occurs at $t=8.6\times 10^{17}\hbox{ sec.}\simeq3\times 10^{10}$ years (that is, about ten times the present age of the universe). This indicates that, in finding the intersection, one should not neglect the curvature term in the general equation of the expanding universe.'' The space curvature term in eq.~[\ref{eq:FL}] certainly is important for the cosmological parameters used here, but the check computation of $t_{\rm eq}$ presented here takes account of that, and it agrees with the shorter time Alpher and Herman (1948a) give in the second paragraph of the paper. I do not understand the longer intersection time in the first paragraph.

 The origin of the matter density Alpher and Herman (1948a) use to find the present radiation temperature  ``of about $5\,^\circ$K'' is suggested by the statement in Alpher and Herman (1949) that
\begin{quotation}
We believe that a determination of the matter density on the basis of only the first few light elements is likely to be in error. Our experience with integrations required to determine the relative abundances of all the elements$^{6,\,7}$ indicates that these computed abundances are critically dependent upon the choice of matter density.
\end{quotation} 
(The references are to Alpher 1948b and Alpher and Herman 1948a.) That is, Alpher and Herman chose parameters to fit solutions to the buildup equation~(\ref{eq:rates}) for a broad range of relative abundances, as opposed to Gamow's consideration of the threshold for element formation. 

This difference of opinion between Gamow and Alpher and Herman is even more explicit in the abstract (Alpher, Herman, and Gamow 1949) for a talk at the 26-27 November 1948 meeting of the American Physical Society. The abstract begins, ``Continuing preliminary work on the neutron-capture theory of the relative abundance of the elements,$^1$ we have investigated in greater detail the following cases:'' (The reference 1 is to Alpher, Bethe, and Gamow 1948.) The first case, on ``the building-up of the elements taking into account neutron decay but not the universal expansion$^2$ gives a better fit to the abundance data than previously reported,$^1$ and requires an initial neutron density of $5\times 10^{-9}$ gm/cm$^3$\,''. This is the density reported in Alpher and Herman (1948b), which likely is reference (2) in this quote, in the abstract stated to be in press. The second case, on ``the building-up of deuterium only, with both neutron decay and universal expansion included$^3$ yields a matter density $\rho = 7.2\times 10^{-3}t^{-3/2}$\,gm/cm$^3$ if the parameters are adjusted to give a relative concentration 0.5 for hydrogen.'' This is the density in equation~(\ref{eq:CGamow}), and the reference (3) in the quote is to Gamow (1948b), still in press at the time  the abstract was written. 

We may conclude that, although it is not stated in Alpher and Herman (1948a), already in this first of the Alpher and Herman papers they had decided to rely on their approach, the first case mentioned in the abstract, rather than Gamow's consideration of the amount of element formation allowed by conditions at the start of buildup. Their approach has a serious problem, however, as discussed next. 

\section{Unexamined assumptions}\label{Sec:unexamined} 

We can take advantage of hindsight to consider examples in this history of a hazard we all face, unexamined assumptions. In the first paper to predict the present temperature of the remnant thermal radiation, the CMB, Alpher and Herman (1948a) present an estimate of the early matter density that may be what they derived from their corrections to Gamow's (1948b) method, or it may be a misprint. However, although the reader is not notified, the matter density used to find the present radiation temperature is much closer to what Alpher and Herman (1949) derive from the fit of equation (\ref{eq:rates}) to ``the relative abundances of all the elements.'' Alpher and Herman's unexamined assumption is that element buildup flows without much hinderance through Alpher's mass gaps at atomic weights 5, 8 and 11. Gamow's (1948a,b) unexamined assumption is that element formation piles up at helium, so as to match his understanding of the cosmic abundance of helium and heavier elements. As discussed in Section~\ref{Sec:AlpherPaper}, the Fermi-Turkevich analysis shows that Gamow was on the right track. 

A second example is the enormous extrapolation of general relativity theory to the scales of cosmology. At the time the theory had passed just one serious test, the precession of the orbit of Mercury. Gamow (1946a) offers a cautionary remark: ``It goes without saying that  one must be very careful in extrapolating the expansion formula to such an early epoch,'' but that is the only comment there or in the 1948 papers reviewed here on the hazard of relying on general relativity theory. It is now seen that general relativity theory passes a demanding network of cosmological tests (Planck Collaboration XIV 2013). On occasion a beautiful theory may reach brilliant success despite initial scarcity of experimental support and confusion along the way. 

Gamow's disregard of the Steady State cosmology (Bondi and Gold 1948; Hoyle 1948) annoyed Fred Hoyle, as one sees in Hoyle's (1950) review of Gamow and Critchfield (1949):
\begin{quotation}
\noindent the authors use a cosmological model in direct conflict with more widely accepted results. The age of the universe in this model is appreciably less than the agreed age of the Galaxy. Moreover it would lead to a temperature of the radiation at present maintained throughout the whole of space much greater than McKellar's determination for some regions within the galaxy.
\end{quotation}
Hoyle is referring to observations of absorption of starlight by the two lowest levels of interstellar CN molecules. McKellar (1941) finds that ``From the intensity ratio of the lines with $K=0$ and $K=1$ a rotational temperature of $2.3^\circ$\,K follows, which has of course only a very restricted meaning.'' McKellar may have had in mind excitation by collisions rather than radiation. The story of how this line of thought was at last examined, and it was realized that Hoyle(1950) was referring to the first measurement of the CMB left from the hot Big Bang, is recalled in Peebles, Page, and Partridge (2009).

Another example is the consistency of estimates of the temperature at the start of element formation, at or near $T_i\simeq10^9$\,K, mentioned in seven of the eleven publications. In Alpher (1948a), ``A temperature above $10^3$~ev, and less than 10~Mev, perhaps of the order of $10^5$~ev (about $10^9\,^\circ$K) appears to be approximately the correct one.'' In Gamow (1948a) it is the ``dissociation energy of deuterium nuclei'', while in Gamow (1948b) and the appendix of Gamow and Critchfield (1949) explanation is confined to the statement that ``the temperature must have been of the order of $10^9\,^\circ$K.'' Alpher (1948b) states that the temperature ``should have been of the order of $10^5$ ev, or about $10^9\,^\circ$K'', ``at which time the neutron capture process could have begun.'' The reasoning behind the latter statement may be indicated in Alpher and Herman (1948b), who state that, when ``the process of neutron capture started, the temperature must have been sufficiently low such that thermal dissociation of nuclei could be neglected. On the other hand, the temperature must have been sufficiently high such that resonance effects in the capture of neutrons would not occur.'' Alpher and Herman (1948b) conclude that a ``temperature of about $10^5\hbox{ ev }\simeq 10^9\,^\circ$K is suggested.''  Alpher and Herman (1949) mention the same considerations, and conclude that  the temperature ``must have been the order of $10^8$--$10^{10}\,^\circ$K.''  Their choice is slightly lower than usual, $T\simeq 0.6\times 10^9$\,K. My understanding of what Gamow (1948a) meant by ``dissociation energy'', the now established picture, is offered in Section~\ref{Sec:Element buildup}. Among the other papers the closest to Gamow's picture is the thermal dissociation mentioned in Alpher and Herman (1948b), but since this paper also mentions lower values of $T_i$ it does not fully capture the idea. I offer the stability of these temperature estimates as an example of the unexamined adoption of a standard parameter value. We have other examples in the history of science. A very relevant one is the unquestioned acceptance in 1948 of a large overestimate of Hubble's constant, $H_o$, by Alpher, Gamow, and Herman on the Big Bang side, and by Bondi, Gold, and Hoyle on the Steady State side. Hoyle (1950) points to the problem for the Big Bang picture, which requires highly special arrangements to make the cosmic expansion time, which scales as $1/H_o$, greater than radioactive decay ages. Gamow (1954) points to the problem for the Steady State picture, which predicts that the mean age of the galaxies is $1/(3H_o)$, and would say that our galaxy is very much older than average yet looks much like neighboring large spiral galaxies. But questioning the large value of $H_o$ was left to observational astronomers.

\section{Summary remarks}\label{Sec:concl}

The problem with the cold Big Bang picture is that one must postulate some special circumstance that prevents element buildup from happening too early, for otherwise the universe would be left without its most abundant element, hydrogen. The hot Big Bang picture does not have this problem because thermal dissociation by the sea of radiation prevents accumulation of deuterons and their conversion to heavier elements until the temperature has dropped to $T_i\simeq 10^9$\,K (as discussed in the last paragraph in Sec.~\ref{Sec:Element buildup}). This leaves a free parameter, the baryon density $n_i$ at $T_i$ that yields the wanted amount of element buildup, but the parameter can be checked. If the wanted $n_i$ (eq.~[\ref{eq:Gni}]) had implied a present universe that is manifestly too hot it would have ended this line of thought. As it happens, the present temperature (which is somewhat lower than the values in eqs.~[\ref{eq:Tf}] that were computed, or could have been computed, in 1948) is just large enough to allow detailed and informative measurements of the radiation intensity spectrum and distribution across the sky. The success of the idea that a theory of element buildup is much simpler in a hot Big Bang exemplifies how, on occasion, simplicity leads in the right direction.

The first paper on element buildup, Alpher, Bethe, and Gamow (1948), is rightly celebrated for its introduction to the established theory of the origin of the light elements. However, this paper does not mention thermal radiation, it does not take note of the inconsistency that motivated the change of thinking to thermal radiation in a hot Big Bang, and it could confuse anyone who cared to check by giving an erroneous matter density that obscures the inconsistency. Alpher (1948a,b), in his thesis and in the version published in {\it The Physical Review}, clearly states the inconsistency. Alpher (1948a,b) and Alpher and Herman (1948b, 1949) point out that the cold Big Bang cosmology used in these papers may be wrong, because it ignores the possible effect of the mass density of radiation on the rate of expansion of the early universe, but they do not take the next step: Show how the radiation might resolve the inconsistency. 

Gamow (1948a) does not acknowledge the inconsistency (though it is spelled out in Gamow 1949), but the evidence before us is that Gamow found the resolution, in ideas that are central to the established $\Lambda$CDM cosmology. The starting idea is that element formation in the early universe would have to have been a rapid non-equilibrium process. Alpher and Herman (1988) and Kragh (1996) trace the history of ideas about the early universe and leave little doubt that Gamow is to be credited for this point. Gamow's key point in 1948 is that element formation in the early universe in a sea of thermal radiation would have started when the temperature had fallen to the point of suppression of thermal dissociation of deuterons (though one would wish he had been clearer about what he meant by dissociation energy). 

By the end of 1948 there was a clear difference of opinion on how best to estimate physical conditions during early element buildup. Alpher and Herman preferred the fit to the general trend of element abundances with atomic weight, while Gamow preferred the constraint from the threshold of accumulation of deuterons. The former underestimates the effect of Alpher's  mass-5 barrier. The latter is on the right path. 

Gamow certainly recognized that the thermal radiation would have remained after element formation, and he pioneered analysis of its effect on structure formation after element formation, but the paper Alpher and Herman (1948a) presents the first estimate of the present radiation temperature. Their numerical value is not significant because it depends on the present baryon density, which was very uncertain, and they compute element buildup along the wrong path. Their lasting contribution is the first example of how the theory of what was happening when the expanding universe was a few minutes old may be tested by measurements some 14\,Gyr later. 

\acknowledgments

I am grateful for the hospitality of the Aspen Center for Physics, where discussions with Jeremy Bernstein and Michael Turner inspired this essay, and to Jeremy Bernstein, Helge Kragh, and Gary Steigman for comments about drafts that improved the essay.


\begin{thebibliography}{99}

\bibitem[Alpher(1948a)]{1948a}  Alpher, R.~A.\ 1948a.   \textit{On the Origin and Relative Abundance of the Elements}. doctoral dissertation, The George Washington University

\bibitem[Alpher(1948)]{1948PhRv...74.1577A} Alpher, R.~A.\ 1948b. A Neutron-Capture Theory of the Formation and Relative Abundance of the Elements. \textit{Physical  Review} \textbf{74}: 1577-1589 

\bibitem[Alpher et al.(1948)]{1948PhRv...73..803A} Alpher, R.~A., Bethe, H., and Gamow, G.\ 1948. The Origin of Chemical Elements. \textit{Physical Review} \textbf{73}: 803-804

\bibitem[Alpher et al.(1953)]{1953PhRv...92.1347A} Alpher, R.~A., Follin, J.~W., and Herman, R.~C.\ 1953. Physical Conditions in the Initial Stages of the Expanding Universe. \textit{Physical Review} \textbf{92}: 1347-1361

\bibitem[Alpher and Herman(1948a)]{1948Natur.162..774A} Alpher, R.~A., and Herman, R.~C.\ 1948a. Evolution of the Universe. \textit{Nature} \textbf{162}: 774-775 

\bibitem[Alpher and Herman(1948b)]{1948PhRv...74.1737A} Alpher, R.~A., and Herman, R.~C.\ 1948b. On the Relative Abundance of the Elements. \textit{Physical Review} \textbf{74}: 1737-1742

\bibitem[Alpherand Herman(1949)]{1949PhRv...75.1089A} Alpher, R.~A., and Herman, R.~C.\ 1949. Remarks on the Evolution of the Expanding Universe. \textit{Physical Review} \textbf{75}: 1089-1095 

\bibitem[Alpherand Herman(1950)]{1950RvMP...22..153A} Alpher, R.~A., and Herman, R.~C.\ 1950. Theory of the Origin and Relative Abundance Distribution of the Elements. \textit{Reviews of Modern Physics} \textbf{22}: 153-212 

\bibitem[Alpher and Herman(1988)]{1988PhT....41h..24A} Alpher, R.~A., and Herman, R.\ 1988. Reflections on early work on `big bang' cosmology. \textit{Physics Today} \textbf{41}: 24-34 

\bibitem[AlpherHernam(2001)]{AlpherHerman} Alpher, R. A. and Herman, R. C. 2001. \textit{Genesis of the Big Bang}. Oxford University Press, Oxford, 214\,pp

\bibitem[Alpher et al.(1948)]{1948PhRv...74.1198A} Alpher, R.~A., Herman, R. C., and Gamow, G.~A.\ 1948. Thermonuclear Reactions in the Expanding Universe. \textit{Physical Review} \textbf{74}: 1198-1199

\bibitem[Alpher et al.(1949)]{1949PhRv...75.324A} Alpher, R.~A., Herman, R. C., and Gamow, G.~A.\ 1949. On the Origin of the Elements. \textit{Physical Review} \textbf{75}: 332

\bibitem[Bethe(1947)]{1947Bethe} Bethe, H. A. \ 1947, \textit{Elementary Nuclear Theory}, Wiley and Sons, New York

\bibitem[Bondi and Gold(1948)]{1948MNRAS.108..252B} Bondi, H., and Gold, T.\ 1948. The Steady-State Theory of the Expanding Universe. \textit{Monthly Notices of the Royal Astronomical Society} \textbf{108}: 252-270

\bibitem[Chandrasekhar and Henrich(1942)]{1942ApJ....95..288C} Chandrasekhar, S., and Henrich, L.~R.\ 1942. An Attempt to Interpret the Relative Abundances of the Elements and Their Isotopes. \textit{Astrophysical Journal} \textbf{95}: 288-298

\bibitem[Gamow(1946a)]{1946PhRv...70..572G} Gamow, G.\ 1946a. Expanding Universe and the Origin of Elements. \textit{Physical Review}  \textbf{70}: 572-573 

\bibitem[Gamow(1946b)]{1946Natur.158..549G} Gamow, G.\ 1946b.  Rotating Universe? \textit{Nature} \textbf{158}: 549 

\bibitem[Gamow(1948)]{1948PhRv...74..505G} Gamow, G.\ 1948a. The Origin of Elements and the Separation of Galaxies. \textit{Physical Review} \textbf{74}: 505-506

\bibitem[Gamow(1948b)]{1948Natur.162..680G} Gamow, G.\ 1948b. The Evolution of the Universe. \textit{Nature} \textbf{162}: 680-682

\bibitem[Gamow(1949)]{1949RvMP...21..367G} Gamow, G.\ 1949. On Relativistic Cosmogony. \textit{Reviews of Modern Physics} \textbf{21}: 367-373

\bibitem[GamowCritchfield(1949)]{junk} Gamow, G. and Critchfield, C. L. 1949. \textit{Theory of Atomic Nucleus and Nuclear Energy-Sources}, Clarenden Press, Oxford, 344\,pp

\bibitem[Hall Harrison(1948)]{1948ApJ...108..310H} Hall Harrison, M.\ 1948. On the Chemical Composition of the Sun from its Internal Constitution. \textit{Astrophysical Journal} \textbf{108}: 310-325

\bibitem[Hayashi(1950)]{1950PThPh...5..224H} Hayashi, C.\ 1950. Proton-Neutron Concentration Ratio in the Expanding Universe at the Stages preceding the Formation of the Elements. \textit{Progress of 
Theoretical Physics} \textbf{5}: 224-235

\bibitem[Hoyle(1948)]{1948MNRAS.108..372H} Hoyle, F.\ 1948. A New Model for the Expanding Universe. \textit{Monthly Notices of the Royal Astronomical Society} \textbf{108}: 372-382

\bibitem[Hoyle(1950)]{1950Obs....70..194.} Hoyle, F. 1950. Nuclear Energy. \textit{The Observatory} \textbf{70}: 194-195 

\bibitem[Hughes1(1946)]{1946PR} Hughes, D. J. 1946a. Radiative Capture Cross Sections for Fast Neutrons. \textit{Physical Review} \textbf{70}: 106-107

\bibitem[Hughes2(1946)]{1946bPR} Hughes, D. J. 1946b. \textit{Manhattan District Declassified Document}, MDDS-27, April 29, 1946

\bibitem[Kragh(1996)]{1996cchd.book.....K} Kragh, H.\ 1996. \textit{Cosmology and Controversy}. 
Princeton University Press, Princeton, NJ, 500\,pp

\bibitem[Lifshitz(1946)]{1946Lifshitz} Lifshitz, E. 1946. On the Gravitational Stability of the Expanding Universe. \textit{Journal of Physics of the USSR} \textbf{10}: 116-129

\bibitem[Mather(2007)]{2007RvMP...79.1331M} Mather, J.~C.\ 2007. Nobel Lecture: From the Big Bang to the Nobel Prize and beyond. \textit{Reviews of Modern Physics} \textbf{79}: 1331-1348

\bibitem[McKellar(1941)]{1941PDAO....7..251M} McKellar, A.\ 1941.  Molecular lines from the lowest states of diatomic molecules composed of atoms probably present in interstellar space. 
\textit{Publications of the Dominion Astrophysical Observatory Victoria} \textbf{7}: 251-272 

\bibitem[Peebles(2013)]{2013arXiv1305.6859P} Peebles, P.~J.~E.\ 2013. Dark Matter. arXiv:1305.6859

\bibitem[Peebles et al.(2009)]{2009fbb..book.....P} Peebles, P.~J.~E., Page, L.~A., Jr., 
and Partridge, R.~B.\ 2009, \textit{Finding the Big Bang}. Cambridge University Press, Cambridge, UK, 571\,pp

\bibitem[Planck Collaboration et al.(2013)]{2013arXiv1303.5076P} Planck Collaboration XIV. 2013. Cosmological Parameters.  arXiv:1303.5076 

\bibitem[Schwarzschild(1946)]{1946ApJ...104..203S} Schwarzschild, M.\ 1946.  On the Helium Content of the Sun. \textit{Astrophysical Journal} \textbf{104}: 203-207

\bibitem[Smoot(2007)]{2007RvMP...79.1349S} Smoot, G.~F.\ 2007. Nobel Lecture: Cosmic microwave background radiation anisotropies: Their discovery and utilization. \textit{Reviews of Modern Physics} \textbf{79}: 1349-1379

\bibitem[Tolman(1931)]{1931PhRv...37.1639T} Tolman, R.~C.\ 1931. On the Problem of the Entropy of the Universe as a Whole. \textit{Physical Review} \textbf{37}: 1639-1660

\bibitem[Tolman(1934)]{1934rtc..book.....T} Tolman, R.~C.\ 1934, 
\textit{Relativity, Thermodynamics, and Cosmology}, Clarendon Press, Oxford, 497\,pp

\end{thebibliography}
\end{document}